# À quoi s'intéressent les enseignants dans les exemples en formation ? Étude de l'utilisation par des stagiaires de ressources basées sur la vidéo

Simon Flandin, Marine Auby & Luc Ria[1]

## Résumé

*Cet article rend compte d'une étude de cas en vidéoformation des enseignants. Considérant le fait que l'exemplification vidéo est une pratique en plein essor bien qu'il y ait peu de consensus dans la littérature concernant sa scénarisation pédagogique, il paraît pertinent de s'intéresser à l'activité déployée par les enseignants dans les dispositifs afin d'en identifier les caractéristiques prometteuses (nature des exemples, types d'association entre eux, modes de documentation, organisation et scénarisation, etc.). Menée dans la théorie sémiologique du cours d'action (Theureau, 2006), notre étude a consisté à confronter six stagiaires, durant deux sessions de 45 minutes, à un dispositif numérique basé sur une pédagogie des trajectoires professionnelles typiques (Durand, 2014 ; Ria & Leblanc, 2011). L'objectif était de mieux comprendre leur utilisation des exemplifications vidéos et les significations qu'elle fait émerger. Les résultats indiquent que les enseignants stagiaires ciblent préférentiellement les ressources relatives aux « règles économiques » ; privilégient les situations de classe, toujours consultées ; s'intéressent particulièrement (même s'ils sont parfois déçus) aux témoignages d'experts ; manifestent initialement une défiance envers les ressources perçues a priori comme théoriques ; s'intéressent progressivement à ces ressources lors de la seconde session. Les résultats sont discutés dans une perspective de conception de dispositifs de vidéoformation des enseignants.*

Cet article s'inscrit dans le domaine de la recherche et de la conception d'environnements d'apprentissage (notamment numériques), et plus précisément dans un champ de pratiques en formation connaissant un essor notable depuis une dizaine d'années, notamment pour la formation des enseignants (Gaudin & Chaliès, 2012), que par commodité nous appelons « *vidéoformation* »[2]. L'intérêt pour la formation de faire visionner à des enseignants des enregistrements vidéos – de leur propre activité ou de celle de pairs – est resté constant à partir du moment où cela a été techniquement réalisable, c'est-à-dire dans les années soixante (Sherin, 2004). Ce constat témoigne d'une motivation persistante des enseignants et de leurs formateurs à utiliser de tels matériaux, alors même que leur intérêt intrinsèque n'est pas toujours bien défini. D'après la revue de littérature de Gaudin et Chaliès (2012), l'utilisation de la vidéo en formation a connu au cours des trois dernières décennies un fort accroissement à l'échelle internationale, et cela pour cinq raisons principales. Les trois premières sont d'ordre pédagogique : elle permettrait un accès à la salle de classe dans l'espace de la formation ; elle faciliterait l'observation d'une diversité de situations d'enseignement et elle aiderait à la connexion entre théorie et pratique. La quatrième est d'ordre institutionnel : elle constituerait un moyen privilégié pour faciliter la mise en œuvre de réformes[3]. La cinquième est d'ordre technique : les progrès exponentiels du numérique et sa relative démocratisation encourageraient fortement l'enregistrement, l'édition, la collection et la diffusion de ressources

---

vidéos. Globalement, les utilisations de la vidéo en formation semblent donc être souvent pédagogiquement, institutionnellement et techniquement justifiées, sinon fondées. Pour autant, elles recouvrent une très grande variété de pratiques, relativement à la nature des savoirs en jeu, l'instrumentation des exemples et des cas « vidéoscopés », ou encore l'accompagnement du processus de généralisation du singulier vers le typique.

Dans cette variété, il semble qu'un consensus se stabilise dans la littérature, selon lequel la vidéo aurait en formation l'intérêt particulier de préserver la richesse « contextuelle » de la pratique en classe ; de pouvoir être cataloguée et combinée avec d'autres supports dans des formats favorisant des enquêtes fécondes sur l'enseignement et l'apprentissage ; et de faciliter l'accès des enseignants à de nouvelles modalités d'intervention (Sherin, 2004). Sur cette même base, des choix très différents et non hiérarchisables peuvent être faits sur le plan de la conception concernant le choix des extraits vidéos. Les pratiques à exemplifier peuvent être choisies pour :

- leur caractère d'exemplarité ou d'innovation, afin d'encourager cette forme de pratique (par exemple dans la plateforme *Zoom sur l'expertise pédagogique*[4]) ;
- leur caractère de « non-exemplarité », afin de faciliter l'identification de leviers de progression (par exemple dans la plateforme *Lesson Observation On-line Platform*[5]) ;
- leur caractère de typicité, afin de faire analyser et déconstruire une pratique ordinaire et poursuivre les deux objectifs précédents (par exemple dans la plateforme *Former à l'intervention en EPS*[6]).

Les recherches s'accordent largement sur le faible potentiel de la seule exemplification vidéo. Aussi de nombreuses études concluent-elles sur la nécessité d'un fort guidage, en plaidant pour une scénarisation « *fermée* » (Santagata & Angelici, 2010 ; Star & Strickland, 2008), que les tâches d'observation et de réflexion soient prescrites et accompagnées par un formateur, consignées sur papier ou intégrées dans un environnement numérique. À notre connaissance, aucune étude n'a seulement pris en considération la possibilité qu'une scénarisation « *ouverte* » puisse être performante. Aussi le cas d'un environnement numérique basé sur l'exemplification vidéo et suffisamment structuré pour permettre des apprentissages via une utilisation – partiellement ou totalement – autonome des enseignants n'a-t-il jamais été étudié. Cela peut s'expliquer par le fait que les dispositifs étudiés consistent la plupart du temps en des entraînements au repérage d'indicateurs considérés comme objectivement pertinents, comme le raisonnement des élèves (*e.g.* Santagata & Guarino, 2011) ; en des explorations collectives d'aspects particuliers de la pratique, sélectionnés et cadrés par un formateur, comme les feedbacks adressés aux élèves (*e.g.* Borko, Koellner, Jacobs & Seago, 2011) ; ou des instructions collectives de problèmes partagés par les enseignants (*e.g.* Miller, 2009). Ces types de formation peuvent difficilement se dispenser de feedbacks réguliers par les formateurs pour faciliter les échanges et lever les « mésinterprétations » des enseignants formés. En revanche, dans une approche développementale de la formation comme la nôtre – décrite dans la partie suivante – on considère que des modalités d'apprentissage prometteuses peuvent émerger de l'activité en vidéoformation en dehors d'un système de taches et de feedbacks.

À ce titre, l'hypothèse selon laquelle les enseignants pourraient acquérir des savoirs pour l'intervention par l'utilisation autonome d'un dispositif vidéo, dans la mesure où ce dispositif est conçu tout ou partie à cette fin, s'avère nouvelle et pertinente. Elle l'est d'autant plus lorsque l'on s'intéresse à tous les objets d'apprentissage, sans les circonscrire à ceux qui sont attendus par les concepteurs mais en étudiant aussi la façon dont les enseignants les spécifient eux-mêmes, au gré des significations produites. Comme le proposent Gaudin et Chaliès (2012), cela demande notamment : d'approfondir la compréhension de ces significations en situation de vidéoformation, en fonction des caractéristiques des exemples vidéos consultés, et de déterminer comment mieux adapter les dispositifs de vidéoformation aux besoins des

---

enseignants. Inscrite dans la théorie sémiologique du cours d'action (Durand, 2008 ; Theureau, 2010), notre étude s'est donné cette double ambition. Pour ce faire, elle s'est intéressée à l'utilisation par des enseignants stagiaires d'un dispositif, le thème n°1 de la plateforme NéoPass@ction[7], construit sur un principe de *variation ordonnée de dispositions à agir en situation* (défini dans la partie suivante) qui offre des exemplifications vidéos de pratiques différentes (typiques, exemplaires, non exemplaires) et (une scénarisation « *ouverte* » basée sur une interface numérique.

## 1. Hypothèses théoriques et technologiques en vidéoformation

Conçu et développé par une équipe pluridisciplinaire à l'Institut français de l'Éducation, le dispositif NéoPass@ction a été publié en ligne en septembre 2010, avec l'ambition d'impulser de nouvelles orientations dans la façon de concevoir des dispositifs d'accompagnement des enseignants. Ses ressources sont prioritairement destinées à la formation des débutants dans des sessions présentielles et/ou à distance, en formation institutionnalisée avec un superviseur universitaire (pour les étudiants de master) ou avec un tuteur (pour les stagiaires), mais aussi potentiellement en autoformation (stagiaires ou néo-titulaires). Il a été conçu selon un processus de spécification progressive des objets et des modalités de formation, par des boucles articulant itérativement recherche et formation. Ce processus mobilise des présupposés théoriques en termes d'aide à la professionnalisation et des analyses de l'activité des enseignants en situation de travail et en situation de formation, conduites dans une approche développementale (Durand, 2008 ; Leblanc et al., 2008 ; Ria & Leblanc, 2011). Le dispositif repose – sans s'y réduire – sur la vidéo comme moyen privilégié de documenter des situations de travail typiquement problématiques pour les enseignants, les dispositions à agir[8] qu'ils mobilisent pour y faire face et l'évolution typique (ou *variation ordonnée*) de ces dispositions à agir dans les trajectoires professionnelles des enseignants. La documentation consiste en l'exemplification vidéo de l'activité dans la classe, « augmentée » par d'autres ressources (points de vue de différents acteurs, analyses textuelles).

Dans cette perspective, il ne s'agit pas d'utiliser la vidéo uniquement pour exemplifier les « bonnes pratiques » du travail enseignant, comme cela est fréquemment le cas en formation, mais pour décrire une succession ordonnée de différentes dispositions à agir que les débutants mobilisent pour faire face à des activités typiquement critiques pour eux-mêmes (des « passages à risque ») mais également cruciales pour la profession (Ria & Leblanc, 2011). Plus largement, la reconstitution des trajectoires professionnelles des enseignants permet d'identifier des situations professionnelles typiques pour eux, c'est-à-dire des épisodes absents chez des enseignants plus expérimentés et présents chez la plupart des débutants et objets d'expériences critiques, voire problématiques. Ces trajectoires sont jalonnées par des « nœuds » à dépasser au cours de leur développement professionnel, et constituent à ce titre un objet de conception pour des formations écologiques accompagnant la maîtrise progressive de ces nœuds (Ria, 2009).

Un long travail d'observatoire (250 heures de données vidéos de classe et d'entretien dans neuf disciplines issues de quatre établissements) a été nécessaire pour identifier et modéliser, à partir de données empiriques conséquentes, ces situations et dispositions à agir présentant un fort degré de typicité du travail enseignant. La scénarisation a ensuite consisté à sélectionner, exemplifier et scénariser les modélisations les plus prometteuses pour en faire des « *artefacts vidéos cibles* » pour la formation des débutants, c'est-à-dire un répertoire de situations de référence indexées à leur travail réel et en prise directe avec leurs préoccupations immédiates ou à venir à court terme. La scénarisation s'est poursuivie par l'organisation de ces artefacts vidéos cibles de manière non linéaire, et sans suivre une planification formelle et détaillée de parcours

---

de navigation qui serait imposée aux formés. Elle visait à offrir des possibilités d'apprentissage des situations professionnelles dans un environnement que l'on peut qualifier de « *propensionnel* » (Jullien, 2009), c'est-à-dire qui multiplie les points de vue et les portes d'entrée dans le travail réel pour favoriser l'émergence de significations. En particulier, la conjonction chez les formés entre une vision égocentrée (voir la situation professionnelle à la première personne) et une vision allocentrée (voir celle-ci à la troisième personne), entre les émotions-intentions vécues et celles imaginées était recherchée car elle favorise le changement de point de vue sur l'activité (Berthoz, 2004) et *in fine* sa transformation effective.

Les six activités typiques de NéoPass@ction sont ainsi conçues comme des activités cibles pour la formation des enseignants dans la mesure où elles font doublement écho à leurs propres expériences (vécues ou anticipées à court terme) et à celles caractéristiques du développement professionnel typique des débutants. Compte tenu de ces liens de proximité et de résonance mis à jour par plusieurs recherches (Flandin & Ria, 2014a, 2014b ; Flandin, Leblanc & Muller, 2015 ; Leblanc, 2012 ; Ria & Leblanc, 2011), les activités cibles peuvent favoriser la perturbation de leurs propres activités en leur permettant d'identifier des actions possibles à la fois efficaces et à la fois compatibles avec leurs dispositions à agir ; et *a contrario* l'invalidation d'actions incompatibles, impossibles. De cette manière, le dispositif de formation ne prescrit ni n'instruit l'enseignant « de l'extérieur » mais encourage le développement de certaines actions plutôt que d'autres. Cette forme d'instrumentation de la vidéo peut être qualifiée de « *pédagogie des trajectoires professionnelles* » (Durand, 2014) : plutôt que prescrire de supposées bonnes pratiques, elle rend intelligible et appropriable, au moyen d'exemples types « augmentés » par leur documentation, une épure du développement du travail enseignant. Elle contribue ainsi à des transformations majorantes des dispositions à agir des formés, dispositions qui deviennent plus efficaces et plus soutenables (Flandin, 2015).

Le premier thème proposé dans NéoPass@ction traite de l'une des situations les plus problématiques du point de vue des novices lors de la prise de fonction et quelle que soit la discipline enseignée : l'entrée en classe et la mise au travail de leurs élèves. Il exemplifie une typologie de six activités typiques, ordonnées des moins maîtrisées aux plus maîtrisées : cette variation est représentée par les vignettes[9] de la frise verticale à gauche de l'interface (figure 1 ci-après). Chaque activité typique est documentée « horizontalement » dans le reste de l'interface, dont la structure est donc fixe mais le contenu différent. Chacune se distingue de la précédente par la modification des composantes de l'expérience de l'enseignant filmé, mais aussi par l'influence sur l'activité collective de la classe, appréhendée du point de vue du climat (plus ou moins propice aux tâches scolaires) et de celui des comportements individuels des élèves (plus ou moins propice aux apprentissages).

Pour montrer et aider à la déconstruction de la complexité du travail enseignant et de sa dynamique de transformation, chaque activité typique est systématiquement documentée par :

- une exemplification vidéo en situation réelle de classe ;
- un extrait d'entretien avec l'enseignant enregistré, en confrontation à l'enregistrement de son activité ;
- mais aussi des témoignages en confrontation à cet enregistrement de novices (extension à la communauté débutante), de chevronnés (extension à la communauté enseignante), de chercheurs (extension à la communauté experte) ;
- et enfin d'aides textuelles à l'analyse réparties sous quatre onglets :
   · « *résumé* » qui résument la situation de référence,
   · « *composantes de l'activité* » qui la décrivent à l'aide des outils de la recherche,
   · « *analyse* » qui proposent une courte analyse par le chercheur-concepteur,
   · « *pistes* » qui renvoient vers un glossaire et des développements périphériques de la plateforme alors en construction ;
- une dernière catégorie intitulée « *Compléments* » apparait pour certaines activités typiques afin de faire figurer des ressources utiles n'entrant pas dans les catégories précitées, par exemple un nouvel enregistrement en classe avec l'enseignant quelques mois plus tard.

---

[9] Dans la description de NéoPass@ction, nous appelons « vignette » un ensemble [image & intitulé].

En résultent 82 montages vidéos d'une durée moyenne de 2 min. 20 sec. (durée minimale de 38 sec.; durée maximale de 9 min. 42 sec.), chacun associé à un titre évocateur s'appuyant sur les *verbatim* des enseignants, au prénom et à une photo de l'enseignant (en classe ou en entretien, selon).

*Figure 1 - Activité typique n°1 de NéoPass@ction avant septembre 2015,
annotée pour le lecteur (Flandin, 2015)*

Les chercheurs-concepteurs ont une visée et des attentes concernant la façon dont il faudrait que les enseignants utilisent le dispositif pour en tirer le meilleur parti. Toutefois, les résultats sur lesquels ils les fondent ont été obtenus dans des situations d'utilisation accompagnées par un chercheur, dont l'intervention s'est caractérisée par une écoute attentive, des questions, des relances et des reformulations, ainsi que de courts dialogues (méthode des verbalisations interruptives). Ces caractéristiques ne sont probablement pas sans conséquence sur les résultats, et l'on peut faire l'hypothèse qu'une utilisation autonome dans un tel environnement produirait des résultats au moins pour partie différents. Or, ce type d'utilisation n'a jamais été étudié dans le programme de recherche-conception de NéoPass@ction (ni ailleurs, comme vu précédemment), alors qu'il correspond à une situation écologiquement très vraisemblable (les enseignants sont nombreux à l'utiliser seuls : on relève un volume de consultation important en dehors des horaires ordinaires de travail et de formation, et sur certaines périodes de

vacances[10]). Il constitue donc un objet pertinent vis-à-vis des objectifs de notre étude : 1) décrire l'utilisation des ressources en rendant compte des significations produites par les enseignants, telles qu'elles témoignent notamment de réflexion, d'intérêt/désintérêt, de validation/invalidation, c'est-à-dire d'activité constructive ; 2) les mettre en relation avec les caractéristiques des exemples vidéos et de leurs modes de consultation, pour identifier les plus prometteuses, et cela en pointant les convergences et les divergences entre activités réelles et activités attendues par les concepteurs.

La visée de l'étude peut être résumée à la question de recherche suivante : comment les ressources de NéoPass@ction sont-elles utilisées par les enseignants stagiaires en autonomie et comment la documentation des exemples types influe-t-elle sur les significations produites au cours de leur activité de vidéoformation ? La théorie sémiologique du cours d'action dans laquelle nos travaux s'inscrivent fournit des outils conceptuels et méthodologiques *ad hoc* pour y répondre. L'activité est conçue comme une activité signe (Theureau, 2006) – extension de l'hypothèse de la pensée-signe émise par Peirce (1931-1935) – qui peut être décrite comme une concaténation de signes (et non comme intuition, logique, flux d'information ou suite d'opérations mentales). Les contenus de conscience d'un acteur sont conçus comme des sémioses enchâssées, c'est-à-dire des unités significatives d'activité se déterminant successivement sous forme de trames (ou chaînes). Ce modèle d'analyse de l'activité repose sur l'hypothèse que celle-ci est descriptible de façon acceptable moyennant des conditions favorables d'observation et d'accès au point de vue de l'acteur. En tant qu'activité-signe, l'activité s'organise de façon temporellement complexe, synchronique et diachronique, d'où la nécessité d'un atelier méthodologique permettant à la fois sa décomposition extrinsèque (données d'observation) et intrinsèque (données d'entretien).

## 2. Démarche et méthode

■ *Participants et contexte*

Les caractéristiques de notre cohorte sont rassemblées dans le tableau 1 (les prénoms ont été changés).

*Tableau 1 - Caractéristiques des participants à la recherche*

| Prénom | Sexe | Âge | Discipline enseignée | Situation professionnelle | Lieu d'intervention |
|--------|------|-----|---------------------|---------------------------|---------------------|
| Anne | F | 23 | Espagnol | Cursus « classique » | Collège X |
| Louis | M | 23 | EPS[11] | Cursus « classique » | Collège X |
| Alban | M | 23 | Chinois | Cursus « classique » | Collège X |
| Aude | F | 25 | Anglais | Cursus « classique » | Lycée Y |
| Émilie | F | 23 | EPS | Cursus « classique » | Lycée Y |
| Charly | M | 56 | Économie et gestion | Seconde carrière | Lycée Y |

Ces six enseignants stagiaires (notés ES, par la suite) suivaient au moment de l'étude une formation universitaire en alternance (stage de pratique accompagnée en établissement scolaire et enseignements à l'université). Ils ont répondu favorablement à la demande des chercheurs concernant la participation à la présente étude selon des modalités contractualisées ensemble.

---

[10] Le relevé statistique des pages visionnées par les utilisateurs indique que près de 44% de la consultation de NéoPass@ction est effectuée entre 17h et 7h.
[11] Éducation Physique et Sportive.

Nous sommes en effet convenus :

- d'une durée cible (contextuellement et minimalement ajustable) de 45 minutes pour les situations d'utilisation de NéoPass@ction, et d'une heure pour les situations d'entretien ;
- de deux sessions (Utilisation de NéoPass@ction + Entretien) réparties sur le premier semestre de l'année scolaire.

Deux types de données ont été recueillies : des données d'observation de l'activité des ES en situation d'utilisation de NéoPass@ction, par l'enregistrement des opérations des ES dans l'interface (capture vidéo de l'écran en temps réel) ; et des données d'entretien de remise en situation (Flandin, Auby & Ria, 2016[12] ; Theureau, 2010) ayant pour support ces traces numériques de l'activité en vidéoformation.

■ ***Données d'observation de la vidéoformation***

L'ES était laissé seul dans la pièce, en situation d'utilisation autonome de NéoPass@ction. L'écran était enregistré en continu à l'aide de l'application *FastStone Capture*[13]*,* qui concatène des captures à un intervalle de temps optimal permettant d'obtenir une impression dynamique très satisfaisante sans requérir trop de mémoire de calcul et de stockage (dix images par seconde). En d'autres termes, l'enregistrement produit était proche de l'écran vu par l'ES au moment de l'utilisation, avec en plus une aide au repérage des actions effectuées à l'aide du pointeur.

À l'issue de la session, l'enregistrement était donc directement exploitable pour l'entretien de remise en situation, effectué quelques minutes après la fin de la session de formation (données qualitatives), et pour son codage et sa description extrinsèque (données quantitatives). Les ES ont tous utilisé la totalité du temps imparti (45 minutes) lors de la première session (parfois en « dépassant » un peu), et entre 32 et 44 minutes lors de la seconde session. À titre indicatif, le temps cumulé de visionnement vidéo possible du thème 1 de NP@ est de 320 minutes, soit quatre heures.

■ ***Données d'entretien de remise en situation (ERS)***

Pour enregistrer à la fois les comportements de l'acteur et les traces d'activité commentées à chaque instant, les ERS ont été enregistrés à l'aide d'un moniteur dupliquant l'écran utilisé par l'ES pour visionner les traces de sa propre activité (figure 2).

*Figure 2 - Extrait d'enregistrement d'un ERS à l'aide de traces numériques de l'activité (Alban)*

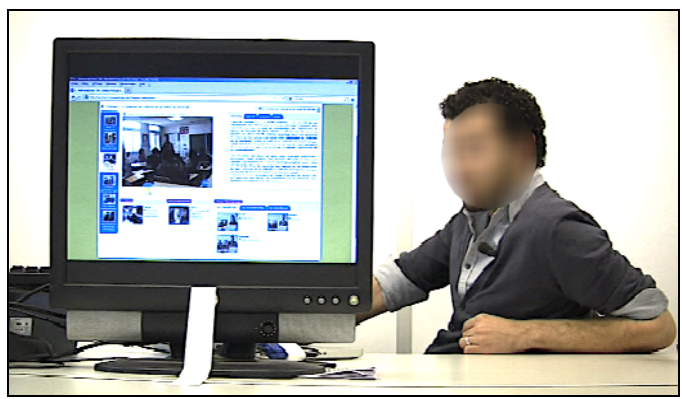

L'enregistrement était diffusé à l'ES sans le son pour deux raisons : (permettre un commentaire simultané limitant la durée totale d'entretien (contrainte) ; encourager la description et l'explicitation. Le chercheur conduisait l'entretien en cherchant à resituer systématiquement l'ES dans son activité de vidéoformation. Or, s'il était nécessaire que l'ES voie les traces visuelles de son action pour la situer et s'y resituer, il n'était pas forcément nécessaire qu'il en entende les traces sonores. En effet, la durée séparant l'utilisation de NéoPass@ction de l'entretien était courte, ce qui facilitait largement le rappel. Les entretiens ont été transcrits *verbatim* avant traitement.

■ ***Traitement des données***

Les données d'enregistrement numérique ont été codées à l'aide d'un protocole quantitatif, sous forme tabulaire et chronologique, ce qui a permis un comptage des occurrences et des durées de consultation des différentes vidéos et catégories de ressources ; puis elles ont fait l'objet d'un traitement statistique descriptif et enfin d'un traitement comparatif systématique : entre les enseignants pour une même session, entre les deux sessions pour un même enseignant et entre les deux sessions pour tous les enseignants.

Les données d'entretien, qualitatives, ont été traitées par l'identification progressive des unités significatives composant l'activité des enseignants ; le repérage rétrospectif de régularités entre ces unités pour une même session ; la comparaison des unités pour un même ES entre ses deux sessions et enfin la comparaison des unités entre les ES. L'étiquetage des unités significatives a résulté du traitement conjoint des données d'observation et d'entretien, vérifié, au besoin, par la consultation des enregistrements originaux. Il s'agit à cette étape de décomposer les actions commentées en unités de signification, considérant le « quasi-isomorphisme » entre les monstrations, récits et commentaires qui sont obtenus de la part de l'acteur en entretien et les significations produites dans l'épisode sur lequel porte le commentaire (Theureau, 2010). L'analyse a permis de décomposer les sessions de vidéoformation en « récits réduits » de l'activité en situation d'utilisation de NéoPass@ction, concaténant en moyenne 50 unités significatives.

## ▌ 3. Résultats

Nous présentons dans cette partie l'analyse relative à l'ensemble des ressources de NéoPass@ction, puis plus spécifiquement aux vidéos de classe, de « *vécu professionnel* » (autoconfrontation de l'enseignant filmé en classe), de *« témoignages de débutants »*, de « *témoignages d'expérimentés* », et enfin de « *témoignages de chercheurs* ». Les statistiques descriptives sont présentées dans deux tableaux en annexe. Le tableau 2 montre la répartition (en nombre et en pourcentage) des vidéos consultées par les ES en fonction de leur activité typique de référence, et pour les deux sessions. Le tableau 3 montre la répartition (en nombre et en pourcentage) des vidéos consultées par les ES en fonction de leur type de ressource, et pour les deux sessions.

■ ***Consultation de l'ensemble des ressources de NéoPass@ction***

Trois résultats principaux peuvent être dégagés de l'analyse de la consultation de l'ensemble des ressources.

Premièrement, tous les ES ont montré une implication soutenue au cours des deux sessions d'utilisation de NéoPass@ction, manifestant régulièrement et de différentes manières leur intérêt pour le thème de l'entrée en classe et de la mise au travail des élèves, pour la centration sur l'activité réelle et son observation vidéo. Toutefois, aucun d'entre eux ne s'est connecté de lui-même à NéoPass@ction postérieurement à cette étude, ce qui semble indiquer que la durabilité de notre dispositif est relativement limitée, par une impression d'épuisement des ressources et/ou un ennui de ne pouvoir échanger avec autrui ; que l'utilisabilité d'un dispositif numérique de formation comme NéoPass@ction ne se situe pas seulement dans l'ergonomie de son interface et dans son accessibilité en ligne, mais implique l'instauration et l'organisation d'espaces de formation dans le quotidien des enseignants, ce qui dans notre étude était organisé et donc facilité par le chercheur.

Deuxièmement, on peut repérer des tendances globales dans l'utilisation des ressources de chacune des six activités typiques dans lesquelles on remarque notamment chez les ES une double tendance : commencer par l'activité typique 1, affichée par défaut et prioriser une ou deux activités typiques-cibles estimées comme les plus proches de leur propre activité. L'activité typique 5 « *Des règles économiques* » s'avère être largement la plus consultée en termes de durée (23% du temps total d'utilisation de NéoPass@ction par tous les ES). La consultation des autres activités typiques s'échelonne de 15,5% pour l'activité typique 3 « *L'écrit-contrôle* » à 8,5% pour l'activité typique 4 « *L'écrit-apprentissage* », la moins consultée en termes de durée. Le temps passé dans ces différentes activités typiques est une donnée significative néanmoins insuffisante : en effet les ES ont globalement tendance à visionner l'intégralité – ou presque – de chaque vidéo consultée, or les vidéos sont de durées très différentes[14] (entre 38 sec. et 9 min. 42 sec.). Il est donc nécessaire de s'intéresser au nombre de vidéos consultées. Au total, 180 vidéos ont été consultées par les ES durant leurs deux sessions d'utilisation de NéoPass@ction, dont 24% dans l'activité typique 5, qui est également la plus consultée en termes d'occurrences, et 7% dans l'activité 6 « *Accueil par la présence physique* », qui est la moins consultée en termes d'occurrences. Les ES ont consulté entre 9 et 22 vidéos par session, avec une moyenne de 15.

Troisièmement, on peut repérer des tendances globales dans l'utilisation des différentes ressources, de façon relativement indépendante de leur activité typique de rattachement. On remarque notamment chez les ES : une forte priorisation des vidéos de classe, qui sont presque toutes vues par tous les ES, et qui représentent en moyenne 26% du temps d'utilisation de NéoPass@ction ; une utilisation à peu près équivalente des autres ressources vidéos (moyenne d'environ 13,5% chacune) à l'exception des « *compléments* » (4%), très peu fournis au moment de l'étude ; et une très nette progression des « *témoignages de chercheurs* » lors de la deuxième session (de 6% à 20%). On constate qu'en termes d'occurrences, la hiérarchie de consultation est très différente : seulement 9% pour les vidéos de « *vécu professionnel* », 17% pour les vidéos de classe et en revanche 27% pour les vidéos de témoignages de chercheur. Concernant la consultation des ressources, les données d'occurrence sont moins pertinentes que les données de durée car 1) certaines ressources ont tendance à être significativement plus courtes et donc à encourager des consultations « en chaine » (c'est le cas des vidéos de chercheurs) et 2) certaines ressources ne proposent qu'une vidéo par activité typique (c'est le cas des vidéos de classe et de « *vécu professionnel* »), et comptent de ce fait moins d'occurrences que les autres. Néanmoins, ces ressources confirment l'évolution de la consultation des vidéos de chercheurs, avec un passage de 13 occurrences en première session à 36 en seconde session, faisant *in fine* des chercheurs la ressource la plus consultée en termes d'occurrences.

■     *Consultation des vidéos de classe*

Les six ES témoignent tous d'un intérêt soutenu pour la vidéo, intérêt que l'on peut préciser comme moyen direct d'observation d'une collection de situations d'enseignement réelles, concrètes, typiques (vidéos de classe) ; comme moyen indirect de réflexion pluri-référencée sur le travail (vidéos d'entretien) ; et comme modalité nouvelle, originale, voire inédite de formation.

Ils réagissent de différentes manières à l'observation des difficultés de leurs pairs ; ils ressentent le plus souvent de l'empathie (en comprenant l'inconfort d'autrui) ; plus rarement de la compassion, se traduisant par l'émergence d'une forte inquiétude de vivre soi-même ces difficultés et plus rarement aussi, de l'agacement, voire un ressentiment envers le collègue, jugé défaillant. Pour autant, aucun des ES n'est resté envahi ou submergé par l'émotion et incapable de distanciation et de réflexion. Le plus souvent, les ES comprennent et adhèrent à la nécessité de s'intéresser aux situations réelles, y compris et surtout problématiques, pour trouver des moyens de les résoudre ou en tout cas de mieux y faire face. À l'inverse, l'observation de situations perçues comme grandement maitrisées peut susciter de l'admiration, mais nous n'avons pas relevé d'inconfort, de formes de découragement ou de dévalorisation personnelle, la « comparaison ascendante »[15] avec un pair conduisant plutôt à une forme d'émulation.

---

[14] Au sujet de la durée des vidéos, les ES n'expriment pas de préférence particulière en termes de format : ils apprécient globalement la relative concision des entretiens et la plus longue durée des vidéos de classe et de « *vécu professionnel* », même si pour la plupart d'entre eux ils « sautent » des passages quand ils ne voient pas l'intérêt de ce qu'ils observent.

L'effet le plus communément observé en début de session d'utilisation de NéoPass@ction a déjà été largement documenté par les études menées en phase de conception initiale. Il s'agit d'une forme de déculpabilisation, de dédramatisation de la difficulté d'abord suscitée par l'observation de situations de classe problématiques :

*« Ben c'est vraiment de se dire que "ben oui en fait..." On n'est pas forcément les seuls à... ben là par exemple c'était le bazar dans la classe, donc se dire "ben moi aussi j'ai fait une erreur, je suis pas la seule à avoir fait cette erreur", que d'autres gens, en l'occurrence ici d'autres débutants, ont eu la même erreur et ont réussi peut-être à trouver une façon nouvelle de résoudre ce problème, quoi. »* (Aude, S1)

Cette déculpabilisation n'encourage pas une déresponsabilisation. Elle participe en revanche d'un gain de soutenabilité du travail qui aide l'ES à mieux temporiser et définir les exigences et les critères de sa propre réussite à court et moyen termes. Elle s'accompagne également d'un sentiment d'appartenance à une communauté débutante en développement, éprouvé aussi via la consultation des entretiens, et renforcé par l'ancrage disciplinaire. La discipline enseignée revêt une importance particulière pour les ES qui sont prioritairement attirés par la leur. La reconnaissance de la discipline accroit le sentiment d'être collègue avec l'enseignant observé, et l'impression de pouvoir véritablement comparer et partager des éléments de la pratique, d'autant plus que la préoccupation la plus récurrente et partagée des ES est l'identification de pistes concrètes pour l'intervention. Cela se retrouve pour les disciplines perçues comme voisines, comme les langues vivantes, et peut constituer un obstacle pour les disciplines perçues comme « éloignées ». Par exemple, les ES d'EPS se reconnaissent peu de proximité avec les autres disciplines principalement parce que l'EPS ne se pratique pas comme elles dans une salle de classe, matériellement organisée par des tables et des chaises. Néanmoins, ceci ne constitue pas toujours un obstacle. Les ES expriment parfois un intérêt fort pour des vidéos relatives à une autre discipline que la leur, ce qui correspond à la visée « transversale » de NéoPass@ction.

■ *Consultation des vidéos de « vécu professionnel »*

Dans NéoPass@ction, le « *vécu professionnel* » est couplé aux situations de classe pour aider les utilisateurs à « *comprendre le sens, l'efficience, les aspects spécifiques et typiques, les ancrages concrets dans l'action, les contraintes liées aux conditions d'enseignement* » (Ria & Leblanc, 2011, p.162). Il s'agit de commentaires situés de l'enseignant autoconfronté à son activité en classe, qui visent à « déplier » le vécu et rendre explicites des éléments importants pour lui dans les situations qui sont non observables, ou alors difficilement et/ou indirectement. Souvent les ES perçoivent le « *vécu professionnel* » de manière différente, plutôt comme une « autocritique » ou une « autojustification » dans laquelle l'enseignant documente moins l'activité qu'il ne tente de la justifier *a posteriori*, ce qui est pourtant précisément ce qu'ont cherché à éviter les concepteurs. Les préoccupations des ES portent majoritairement sur l'identification de solutions concrètes pour leur propre activité, ce qu'ils ne trouvent que très rarement « directement » au premier visionnage des vidéos de « *vécu professionnel* ». Néanmoins, au cours des sessions, les ES font souvent l'expérience d'au moins un vécu professionnel qu'ils jugent utile. C'est le cas d'Alban lorsqu'il consulte celui de Cécile dans l'activité typique 6 *« Accueil par la présence physique »* :

*« Ben je l'ai trouvée, j'ai trouvé son retour assez pertinent… Ce qui m'a vraiment marqué de cette vidéo de retour ben c'est la capacité qu'elle a finalement à gérer une classe qui peut très vite être difficile, et elle donne des techniques très simples. Donc la voix, le regard, c'est des*

---

*choses qu'on peut tous appliquer et qui marchent… Directement applicables, sans avoir besoin de prendre le temps à la maison de construire des fiches, de construire des aménagements à nos séquences, des choses comme ça. C'est des règles qui sont accessibles et du jour au lendemain quoi.* » (Alban, S1)

Dans cet extrait, Alban indique avoir satisfait ses attentes relatives à l'identification d'actions simples, efficaces et accessibles, « *directement applicables* » dans sa classe sans devoir augmenter sa charge de travail déjà conséquente. Il apprécie dans cette vidéo de vécu professionnel la formulation de liens explicites entre les préoccupations et les actions de Cécile, qu'il nomme successivement « techniques » et « règles ». Une appréhension globale du corpus de données montre que lorsque les ES rendent compte des chaines de significations les conduisant à l'apprentissage de nouvelles actions, ils ont parfois du mal à se remémorer précisément les interactions dans NéoPass@ction ayant contribué aux différentes significations produites.

■ *Consultation des vidéos de « témoignages de débutants »*

Les « *témoignages de débutants* » sont des lectures croisées des situations de référence par des membres de la communauté novice, qui les analysent à l'aune de leurs propres expériences dans des situations similaires ou parentes. Ils permettent de créer une sorte de collectif virtuel de novices dans lequel l'ES peut accéder à différents points de vue et affiner le sien (par validation et/ou invalidation des modalités d'action). Il ne s'agit donc pas d'analyses longuement réfléchies et conceptuellement très étayées, mais de réactions « à chaud » explicitant des difficultés partagées et traduisant des préoccupations vives et les savoirs qu'elles sélectionnent. Même si ces principes ne sont pas toujours saisis au départ par les ES, ils finissent tous par les vivre peu ou prou comme tels au cours de la session.

L'analyse montre que les significations produites par les ES au cours de leur consultation des témoignages de débutants s'organisent en trois domaines, qui peuvent être formulés ainsi : 1) il est normal de commettre des erreurs, de rencontrer des problèmes, et de chercher à les résoudre ; 2) enseigner s'apprend et la mise en commun des stratégies développées par les débutants qui sont parvenus à les dépasser est un levier porteur ; 3) le développement d'une efficacité satisfaisante est possible à court terme. D'emblée et d'un point de vue extérieur, ces significations peuvent sembler triviales : en réalité souvent, comme l'exprime Louis, « *même si on le sait, on a besoin de le voir. On a vraiment besoin de le voir.* » Cela signifie que même si « dans l'absolu », les ES savent qu'en tant que novices il est très probable, et peu condamnable, que leur efficacité soit moyenne, le constater de manière factuelle dans les classes et dans les entretiens renforce considérablement ce savoir. De plus, comme l'exprime Aude dans l'extrait suivant, les ES n'ont pratiquement aucune occasion de faire ce constat :

« *Des débutants aussi, en tout cas ce que moi je ressens, c'est que même si on a le tuteur qu'on peut aller voir en cours et qui vient nous voir en cours, on a quand même des moments où on a envie de savoir "est-ce que ce que je fais c'est bien ?" et on n'a pas de retour, vraiment, de comment ça se passe ailleurs. Et c'est ce qui est bien dans les vidéos témoignages des débutants, c'est qu'on a d'autres façons de faire, où l'on voit comment d'autres... Ouais on se sent moins seul en fait. On peut comprendre d'autres pratiques.* » (Aude, S1)

Aude souligne ici l'intérêt de prendre connaissance de la réalité du terrain chez les pairs mais aussi la possibilité de se situer, vis-à-vis de l'efficacité « normale » des débutants : « *est-ce que ce que je fais c'est bien ?* » se demande-t-elle. À ce titre, les témoignages de débutants nourrissent une quête de légitimation des ES.

■ *Consultation des vidéos de « témoignages d'expérimentés »*

Les témoignages des expérimentés sont globalement privilégiés par les ES, surtout au cours de la première session. Ceux-ci évoquent l'intérêt d'anticiper des scénarios non encore rencontrés et de « capitaliser » par procuration le vécu des enseignants notamment les plus expérimentés, ce qui correspond à deux hypothèses de conception de NéoPass@ction. Les expérimentés sont perçus comme les acteurs les plus à même de délivrer des solutions concrètes aux problèmes de métier, ce qui est prioritairement recherché par les ES dans NéoPass@ction. C'est pourquoi l'âge d'un enseignant, estimé via l'image qui accompagne la vidéo de son témoignage, constitue parfois une affordance invitant à la consultation :

« *J'ai choisi Jacinthe plutôt qu'Anne-Laure parce que j'ai vu une différence d'âge en fait entre Anne-Laure et Jacinthe… Et je me suis dit effectivement qu'Anne-Laure a certainement pas autant d'années de métier que Jacinthe et j'ai voulu voir ce que quelqu'un de très expérimenté pouvait proposer comme analyse.* » (Alban, S1)

Le principe de capitalisation du vécu des expérimentés rencontre facilement les préoccupations des ES. En revanche, peu s'interrogent sur les conditions de transmissibilité, ou de transférabilité des actions montrées et/ou commentées dans NéoPass@ction. Quand de tels raisonnements apparaissent, ils questionnent surtout la possibilité d'adapter à sa discipline des actions ancrées dans une discipline différente. En outre, bien que les actions perçues comme efficaces soient souvent attribuées à ce qu'ils conçoivent comme un « *style* », à un « *talent* » personnel, l'opportunité d'appliquer *in extenso* à leur activité débutante des actions relevant d'une activité experte semble très peu questionnée. Les vidéos de témoignages d'enseignants plus expérimentés peuvent donc offrir des opportunités concrètes d'appréhension de modalités d'action *a priori* efficaces ; la principale limite est un risque de mécompréhension par les ES des enjeux et conditions de leur transposition en classe.

■ ***Consultation des vidéos de « témoignages de chercheurs »***

Lors de la première session, les ES consacrent en moyenne 6% de leur temps aux témoignages de chercheurs (entre 0 et 14% selon les ES), et en moyenne 20% lors de la deuxième session (entre 5 et 38% selon les ES). Si ces vidéos ne sont pas spontanément estimées par les ES comme ayant un intérêt majeur dans NéoPass@ction, leur perception change sensiblement pour chacun d'entre eux dans le temps : à mesure qu'ils les consultent, ils leur confèrent progressivement une plus grande utilité. Ces ressources finissent par être les plus consultées par les ES durant les deux sessions.

Lors de la première session, certains ES évoquent une impression de ressources « conclusives », qui viendraient faire la synthèse ou le bilan de la formation, impression qui expliquerait en partie leur évincement au cours de la première session. Les ES peuvent aussi penser pouvoir se dispenser des chercheurs s'ils ont « *déjà compris* » à l'aide des autres ressources. De plus, certains sont rebutés par ce qu'ils perçoivent comme un vocabulaire scientifique inaccessible : présents dans les intitulés des vidéos, ce vocabulaire peut dissuader l'ES de les consulter. C'est plusieurs fois le cas pour Aude avec des mots comme « *doxa* » ou « *dynamique de transformation* ».

La tendance la plus prégnante parmi les ES est bien celle d'une proximité avec les pairs et d'une distance avec les chercheurs. Cette distance témoigne : d'un « surplombement » des pratiques perçu de la part les chercheurs et d'une « défiance » à l'égard de la théorie. En pleine construction de leur identité professionnelle, les ES ont tendance à rechercher la légitimation et les conseils de leur nouvelle communauté d'appartenance, et à rejeter *a priori* les analyses théoriques telles que les témoignages de chercheurs dans NéoPass@ction :

« *En fait j'ai l'impression que les chercheurs je les sentirais peut-être plus loin de la réalité. Vu qu'ils sont chercheurs, et que leur but c'est de chercher, ils sont pas vraiment en classe, devant leur classe, à affronter... tu vois ce que je veux dire... ils sont vraiment plus éloignés.* » (Aude, S2)

Dans cet extrait, Aude exprime cette défiance envers la capacité des chercheurs à apporter un éclairage concret sur les problèmes professionnels du métier. S'ils n'en ont pas moins de légitimité à les étudier – puisque c'est précisément cela, leur métier – donner « en extériorité » des conseils pour les résoudre semble moins accepté. Cependant, les ES s'aperçoivent progressivement que d'une part les témoignages de chercheurs ne sont pas aussi surplombants qu'ils le craignaient, et que d'autre part ils sont le plus souvent ancrés dans la pratique, voire dans la situation de classe de référence, et non déconnectés des réalités professionnelles.

Les « *témoignages de chercheurs* » sont les seules ressources à connaître une évolution significative, au cours du temps, de leur consultation par les ES. Cette évolution est principalement due à un décalage entre la faible utilité perçue par les ES avant la consultation des ressources, et l'utilité réelle estimée et renforcée positivement avec leur consultation. Elle montre que la contribution des chercheurs peut être appréciée par les ES si elle consiste à déconstruire les situations professionnelles proposées, dont la complexité peut leur échapper.

## Conclusion

Les résultats de notre étude fournissent plusieurs clés de compréhension de l'activité des ES en situation de vidéoformation, des significations qu'ils produisent et de l'intérêt qu'ils portent aux exemples consultés selon leur nature.

Ils montrent notamment que cet intérêt se déploie dans des temporalités différentes selon la nature des ressources. Les ES privilégient en effet d'emblée les situations de classe, qui favorisent très rapidement le positionnement de leur activité vis-à-vis de celles exemplifiées sur NéoPass@ction. Ce positionnement a tendance à s'accompagner d'une forme de déculpabilisation qui est intéressante en formation car elle aide les ES à assumer leurs difficultés et à chercher à les dépasser plutôt qu'à les masquer. Même s'ils s'avèrent parfois en décalage avec leurs attentes une fois consultés, les témoignages d'experts suscitent aussi assez spontanément l'intérêt des ES, ancré dans des préoccupations relatives à l'identification immédiate de solutions efficaces et directement employables dont ils pensent que les expérimentés sont détenteurs – mais qu'ils trouvent rarement. *A contrario*, les ES manifestent initialement une défiance envers les ressources perçues *a priori* comme théoriques (témoignages de chercheurs et textes d'accompagnement) et ne s'intéressent que progressivement à ces ressources lors de la seconde session. Ce résultat montre que faire l'expérience d'aides à l'analyse en phase avec leurs préoccupations du moment, leur permettant de mieux composer avec des problèmes rencontrés ou susceptibles d'être rencontrés dans la pratique, encourage une « conciliation » (ou « réconciliation ») des ES avec les aides conceptuelles, ou plus simplement leur appropriation. Le principe de conception « propensionnel » de NéoPass@ction, concrétisé par des ressources multipliant les points d'entrée dans l'activité, le décentrement de point de vue, la résonance et la comparaison, semble fonctionnel dans le cadre d'une utilisation autonome des ressources. Il favorise de manière régulière l'engagement des ES en créant des points d'intérêt, mais trouve sa limite dans la capacité réduite qu'ont les ES à mettre en relation ces différents points dans une problématique de formation.

Les résultats indiquent également que les ES ciblent préférentiellement les ressources relatives à l'activité typique n°5 « *règles économiques* », qui semble être la plus porteuse de leur point de vue, c'est-à-dire à la fois proche de leur propre activité et prometteuse en termes d'apprentissage. Cette « auto-situation » dans la trajectoire générique exemplifiée par NéoPass@ction semble être un phénomène particulièrement déterminant en situation d'utilisation autonome. En effet, lorsque le décalage entre l'activité cible et l'activité vécue par les débutants est optimal, les ES expriment un fort intérêt. Celui-ci se traduit par une enquête dont les caractéristiques sont une attention soutenue, une valorisation des solutions investiguées et une conscientisation de leur capacité à apprendre et à transformer leur activité. Toutefois, cette enquête reste majoritairement polarisée sur le rapport entre activité vécue et activité en train d'être visionnée. La navigation autonome ne semble pas permettre aux ES de déconstruire les activités typiques de façon à comprendre la trajectoire dans laquelle elles s'inscrivent, et ainsi produire des significations non seulement sur ces activités mais aussi sur leur mode de développement. Le principe de variation ordonnée des dispositions à agir sur lequel est basé

NéoPass@ction favorise donc l'engagement des ES par la projection de leur activité dans des activités semblables et peut contribuer à ce qu'ils apprennent des actions nouvelles, plus pertinentes que celles dont ils disposent déjà, mais accessibles ; toutefois, ce principe ne se dispense pas d'aides complémentaires pour être exploité de façon optimale en formation.

Lorsqu'elle est basée sur des postulats développementaux, la conception de dispositifs vidéos nécessite donc de garantir un air de famille entre ce qui est vu en vidéo et ce qui est vécu en pratique par l'enseignant pour générer des résonances entre les deux, tout en garantissant une différence suffisante pour amorcer des enquêtes. Les limites au potentiel de la vidéoformation autonome pointées par cette étude indiquent aussi, en creux, la plus-value déterminante que peut apporter l'intervention d'un formateur. Pour tirer le meilleur parti des exemples vidéos, celui-ci devrait notamment aider les ES à déconstruire les différentes activités observées (savoirs, techniques, stratégies mobilisés par les enseignants et effets sur l'activité des élèves et donc sur leurs apprentissages) pour mieux comprendre pourquoi et comment en changer. Il pourrait aussi contribuer à des domaines de significations absents de la vidéoformation autonome, comme l'élaboration d'une problématique de navigation (Flandin & Ria, 2014), l'explicitation de ce que le visionnement de films suscite comme réactions chez eux (Leblanc, 2012), l'identification de leurs propres croyances et modes d'agir à travers ceux des autres (Ria & Leblanc, 2011), à imaginer des nouvelles manières d'agir ou encore à élaborer des conflits de normes susceptibles de déboucher sur des renormalisations (Lussi Borer & Muller, 2014). Si, comme nous l'avons expliqué plus en détail dans un autre texte (Flandin, 2015), nous ne plaidons pas pour une formation des enseignants à l'analyse de l'activité, les conclusions de notre étude laissent en revanche à penser que pour tirer le meilleur parti d'exemples vidéos en formation, les formateurs doivent être formés à ses théories et méthodes, connaître les effets qu'ont tendance à générer différentes utilisations, et enfin disposer d'un modèle de développement professionnel des enseignants (Ria, 2009, 2012).

## Références

THEUREAU J. (2010), « Les entretiens d'autoconfrontation et de remise en situation par les traces matérielles et le programme de recherche "cours d'action" », *Revue d'anthropologie des connaissances*, vol. 42, n°2, p.287-322.

## Annexes

*Tableau 2 - Répartition des vidéos consultées par les ES en fonction des activités typiques (nombre et pourcentage) et des deux sessions*

| | | Aude | | Anne | | Alban | | Charly | | Émilie | | Louis | |
|---|---|---|---|---|---|---|---|---|---|---|---|---|---|
| | Activité typique | Nb | % | Nb | % | Nb | % | Nb | % | Nb | % | Nb | % |
| Session 1 | A1 | 5 | 24% | 6 | 27% | 9 | 60% | 1 | 7% | 0 | 0% | 5 | 42% |
| | A2 | 6 | 29% | 7 | 32% | 0 | 0% | 2 | 13% | 0 | 0% | 0 | 0% |
| | A3 | 4 | 19% | 9 | 41% | 0 | 0% | 0 | 0% | 0 | 0% | 2 | 17% |
| | A4 | 0 | 0% | 0 | 0% | 3 | 20% | 3 | 20% | 0 | 0% | 2 | 17% |
| | A5 | 4 | 19% | 0 | 0% | 0 | 0% | 9 | 60% | 7 | 78% | 1 | 8% |
| | A6 | 2 | 10% | 0 | 0% | 3 | 20% | 0 | 0% | 2 | 22% | 2 | 17% |
| Session 2 | A1 | 2 | 15% | 0 | 0% | 0 | 0% | 2 | 11% | 0 | 0% | 2 | 18% |
| | A2 | 1 | 8% | 0 | 0% | 0 | 0% | 0 | 0% | 11 | 79% | 0 | 0% |
| | A3 | 5 | 38% | 0 | 0% | 14 | 100% | 1 | 6% | 3 | 21% | 0 | 0% |
| | A4 | 3 | 23% | 7 | 44% | 0 | 0% | 8 | 44% | 0 | 0% | 0 | 0% |
| | A5 | 2 | 15% | 8 | 50% | 0 | 0% | 7 | 39% | 0 | 0% | 6 | 55% |
| | A6 | 0 | 0% | 1 | 6% | 0 | 0% | 0 | 0% | 0 | 0% | 3 | 27% |

*Tableau 3 - Répartition des vidéos consultées par les ES en fonction des types de ressources (nombre et pourcentage) et de la session*

| | | Aude | | Anne | | Alban | | Charly | | Émilie | | Louis | |
|---|---|---|---|---|---|---|---|---|---|---|---|---|---|
| | Ressource | Nb | % | Nb | % | Nb | % | Nb | % | Nb | % | Nb | % |
| Session 1 | Classe | 4 | 19% | 2 | 9% | 4 | 27% | 5 | 33% | 2 | 22% | 3 | 25% |
| | Vécu | 3 | 14% | 2 | 9% | 2 | 13% | 1 | 7% | 2 | 22% | 2 | 17% |
| | Complém. | 0 | 0% | 0 | 0% | 1 | 7% | 0 | 0% | 2 | 22% | 0 | 0% |
| | Débutants | 7 | 33% | 4 | 18% | 1 | 7% | 0 | 0% | 3 | 33% | 1 | 8% |
| | Experts | 7 | 33% | 9 | 41% | 7 | 47% | 4 | 27% | 0 | 0% | 3 | 25% |
| | Chercheurs | 0 | 0% | 5 | 23% | 0 | 0% | 5 | 33% | 0 | 0% | 3 | 25% |
| Session 2 | Classe | 1 | 5% | 2 | 13% | 1 | 7% | 3 | 17% | 2 | 14% | 2 | 18% |
| | Vécu | 0 | 0% | 1 | 6% | 1 | 7% | 1 | 6% | 1 | 7% | 1 | 9% |
| | Complém. | 1 | 5% | 0 | 0% | 0 | 0% | 0 | 0% | 0 | 0% | 0 | 0% |
| | Débutants | 0 | 0% | 5 | 31% | 3 | 21% | 4 | 22% | 5 | 36% | 0 | 0% |
| | Experts | 0 | 0% | 3 | 19% | 4 | 29% | 4 | 22% | 5 | 36% | 0 | 0% |
| | Chercheurs | 11 | 85% | 5 | 31% | 5 | 36% | 6 | 33% | 1 | 7% | 8 | 73% |